# DESIGN OF THE JLC/NLC RDDS STRUCTURE USING PARALLEL EIGENSOLVER OMEGA3P *

Z. Li, N.T. Folwell, K. Ko, R.J. Loewen, E.W. Lundahl, B. McCandless, R.H. Miller, R.D. Ruth, M.D. Starkey, Y. Sun, J.W. Wang, SLAC; T. Higo, KEK

*Abstract*

The complexity of the Round Damped Detuned Structue (RDDS) for the JLC/NLC main linac is driven by the considerations of rf efficiency and dipole wakefield suppression. As a time and cost saving measure for the JLC/NLC, the dimensions of the 3D RDDS cell are being determined through computer modeling to within fabrication precision so that no tuning may be needed once the structures are assembled. The tolerances on the frequency errors for the RDDS structure are about one MHz for the fundamental mode and a few MHz for the dipole modes. At the X-band frequency, these correspond to errors of a micron level on the major cell dimensions. Such a level of resolution requires highly accurate field solvers and vast amount of computer resources. A parallel finite-element eigensolver Omega3P was developed at SLAC that runs on massively parallel computers such as the Cray T3E at NERSC. The code was applied in the design of the RDDS cell dimensions that are accurate to within fabrication precision. We will present the numerical approach of using these codes to determine the RDDS dimensions and compare the numerical predictions with the cold test measurements on RDDS prototypes that are diamond-turned using these dimensions.

## 1 INTRODUCTION

The 1 TeV JLC/NLC[1,2] collider consists of ten thousand X-band accelerator structures. Each structure is made up of 206 cells with dimensions tailored from cell to cell to detune the dipole modes frequencies to suppress the dipole wakefields. The cells are optimized to have a round profile for higher rf efficiency. With the manifolds and slots added to damp the wakefields, the structure cells become fully 3 dimensional and complex, as shown in Fig. 1. A time and cost saving approach for the structure design is to determine the 3D RDDS dimensions through computer modeling and to machine the cells using high precision diamond turning machines. With the advanced modeling and machining tools, the structure can be designed and manufactured with a greater accuracy such that no tuning will be need once the structure is assembled. The frequency tolerance on the fundamental mode of the X-band structure is about 1~MHz in order to maintain a better than 98% acceleration efficiency in the structure. At the frequency of 11.424~GHz, this tolerance corresponds to one micron in the major dimensions of the cell. To model the RDDS with such an accuracy, vast computer resources and accurate field solvers are required. With the newly developed finite-element parallel code Omega3P running on massive parallel supercomputers, such an accuracy is readily achievable. In this paper, we will present the numerical design of the RDDS structure using the Omega3P code. We will compare the numerical results with the rf measurements on the high precision machined cells.

## 2 RDDS STRUCTURE

The round damped-detuned structure (RDDS)[3] was designed to suppress wakefields in the JLC/NLC linacs. The RDDS consists of 206 cells connected via slot openings to four pumping manifolds that run the length of an accelerator section. The dimensions of the cells are chosen such that the deflecting modes are detuned in a prescribed manner such that the wakefields are decohered and reduced in magnitude. The coupling slots provide the conduit by which the wakefields within the structure can escape out to the manifold to be absorbed externally, thereby further decreasing the wakefields. The fundamental frequency of the RDDS needs to be determined within 1~part of 10,000, which corresponds to an accuracy in the 3D cell dimensions of one micron. This complex shape can only be modeled accurately by conformal meshes on unstructured grids.

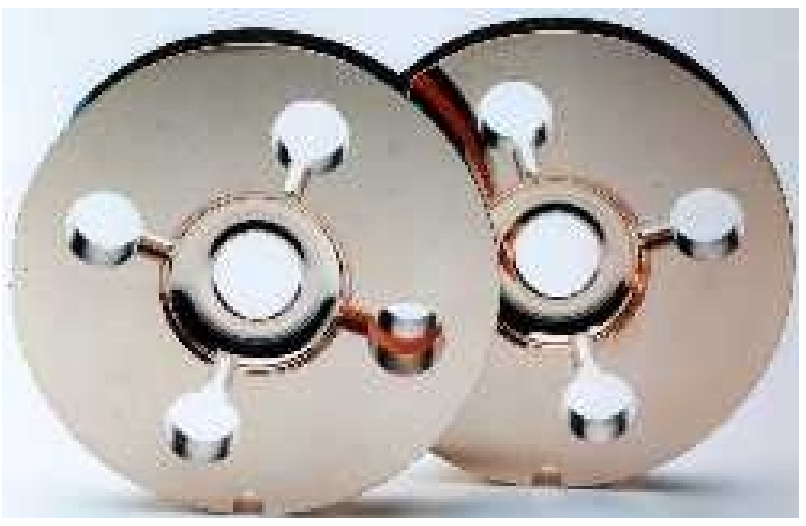

Figure 1. Round damped-detuned structure: cell profile optimized for rf efficiency, dipole modes detuned and damped to suppress long-range wakefield.

## 3 OMEGA3P

Omega3P[4,5], a thrust effort in the DOE Grand Challenge, is a new 3D eigensolver designed to model large, complex RF cavities with unprecedented accuracy and speed by employing advanced eigensolver algorithms on massively parallel supercomputing platforms. The code uses quadratic finite element formulations for field interpolation, uses unstructured mesh to model complex geometries. The code takes advantage of existing parallel libraries such as ParMETIS[6] for mesh partitioning and load balancing, and AZTEC[7] for scalable matrix-vector operations. The unstructured mesh is generated using existing packages such as SIMAIL[8] and CUBIT[9]. The mesh generator can take geometry inputs from solid models obtained with popular CAD tools such as SolidEdge and EMS. Omega3P is written in C++ programming language and currently runs on supercomputing platforms such as SGI T3E, IBM SP2, and Linux clusters. The code also runs on single processor workstations.

*Work supported by the DOE, contract DE-AC03-76SF00515.

With the parallel code Omega3P, the computational domain is divided into smaller subdomains via domain decomposition with each subdomain fitted into one of the processors for computation. This allows to model large complex problems with much higher grid resolutions within reasonable run time, far beyond the computational resource that can be provided by single processor workstations. It provides a tool that is essential for modeling complex rf components such as the RDDS structure.

### *3.1 Omega3P convergence*

The convergence of Omega3P on modeling the RDDS1 cell is as shown in Fig. 2. The color-coded mesh shows the domain decomposition for the 16-processor parallel computation. The mesh was generated using CUBIT. The frequency converges as the fourth power of the mesh size. To reach a better than 1 MHz accuracy for the RDDS cell, the run time is less than 20 min on 16 processors.

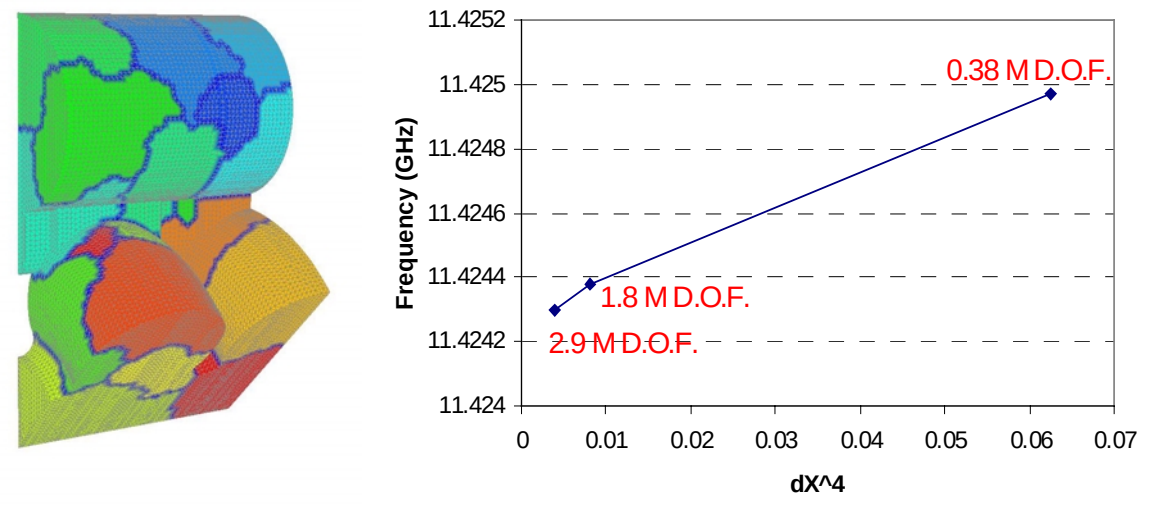

Figure 2. Paralell modeling of the RDDS1 cell. Left) domain decomposition, each color assigned to one processor; right) convergence of the Omega3P eigen solver, converges as the fourth power of the mesh size.

### *3.2 Comparison with cold test cells*

Three test cells were calculated using Omega3P on NERSC SGI/CRAY T3E with about 3 million degrees of freedom. The dimensions of the test cells were from the rough machine table. These cells were diamond tuning machined. The cell profiles were measured using a high precision mechanical CMM machine. The profile error and skin depth effect were corrected on the measured frequencies in order to compare with the numerical results.

Table 1. Comparison of numerical result using Omega3P and the cold test measurement on diamond turning machined cells.

| Cell | Numerical(MHz) | Meas.(MHz) | Diff.(MHz) |
|---|---|---|---|
| 001 | 11420.57 | 11420.3 | 0.27 |
| 102 | 11420.35 | 11420.4 | -0.05 |
| 203 | 11420.09 | 11419.7 | 0.39 |

Omega3P results agree with the cold test measurement within half a MHz. The dimensions determined by using Omega3P has reached an accuracy well within the manufacturing tolerance. These benchmark results give us the confidence of determining the RDDS dimensions numerically using Omega3P. As a result, the number of cold test cells for the RDDS1 structure was reduced, minimizing the cost and turn around time of structure R&D.

## 4 DETERMINE RDDS1 STRUCTURE MACHINE TABLE USING OMEGA3P RESULTS

The RDDS1 structure machine table was generated based on the numerical results of the Omega3P calculation. The dimensions of seven cells along the structure were calculated using Omega3P, running on NERSC SGI/CRAY T3E, with up to 3 million degrees of freedom for the numerical model. The third-order Spline function interpolation was used to obtain the dimensions of the rest of the cells. Since the code models only perfect conducting materials, the skin depth effect of 0.7 MHz for copper was corrected for the final machine table.

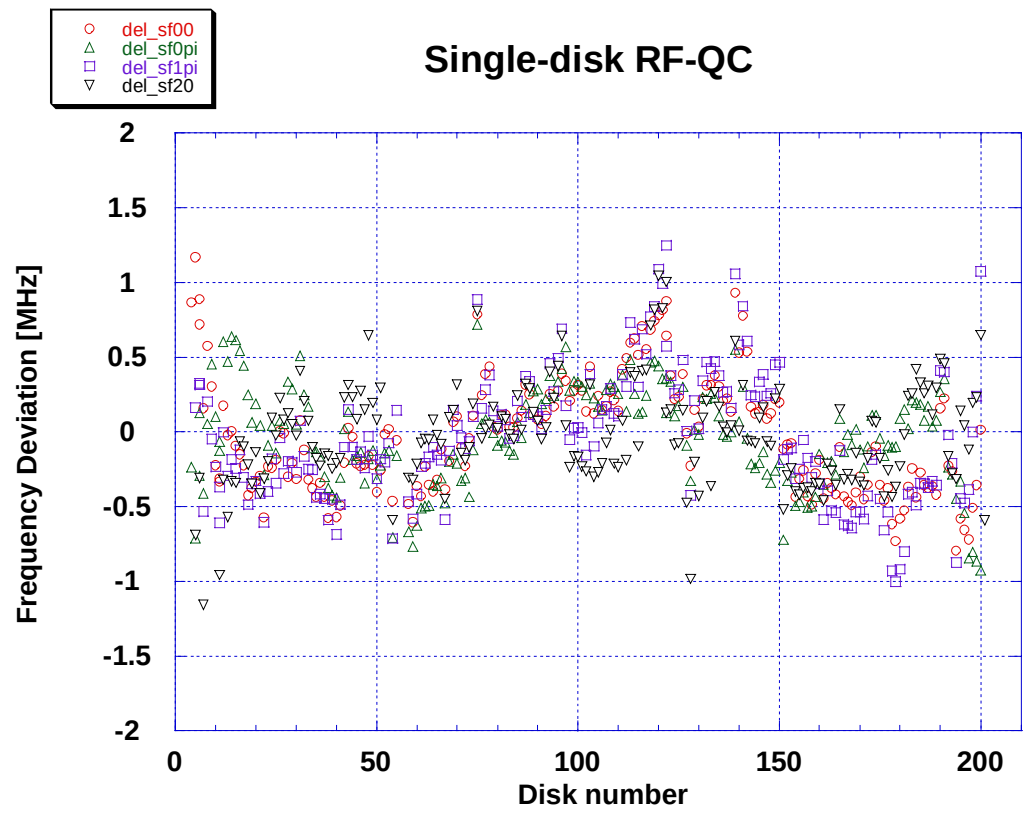

Figure 3. Single cell RF QC of the RDDS1 cells. Shown are the deviations of the fundamental and dipole modes at zero and PI phase advances.

The 206 RDDS1 structure were diamond tuning machined[10]. The frequency of zero and PI phase advance of the fundamental and dipole modes were measured for each disk. The deviations of the frequencies from a smooth curve fitting of the corresponding mode are as shown in Fig. 3. The rms values for all of the four modes are better than 0.5 MHz. We are confident at this point that both the numerical predictions and high precision machining of the 3D RDDS cell dimensions have achieved an accuracy better than $10^{-4}$ in frequency or 1 micron in major cell dimensions, which is well within the NLC design tolerance.

## 5 RF PULSE HEATING

Damping slots perturb current of accelerating mode, which causes additional wall loss and reduces the Q of the accelerating mode. This wall loss is concentrated in a small region around the slot opening, as shown in Fig. 4. The power density, if high, can cause large temperature rise during the rf pulse. It is shown in ref[11] that a temperature rise of $120^0$C can result in structural damage on the copper surface. It is important to evaluate and minimize, if necessary, this pulse heating. Since the high power is distributed within a thin region, a good mesh resolution is need to calculate the power density accurately, to which Omega3P is an ideal tool. The temperature rise on the copper surface can be estimated using the following formula[12]

$$\Delta T = R_S \left|H_{wall}\right|^2 \sqrt{\frac{t_{pulse}}{\pi \rho c k}}$$

where $R_s=1/\sigma\delta$ is the surface resistance, $H_{wall}$ is the magnetic field on the wall (proportional to the wall current), $T_{pulse}$ is the rf pulse length, ρ is the density of copper, $c$ is the specific heat of copper, and $k$ is the thermal conductivity.

At an average gradient of 70 MV/m, the rf pulse heating temperature ris for the RDDS1 structure is from $25^0$C to $55^0$C depending on the location along the structure as shown in Fig. 4. As a comparison, the temperature rise for the detuned RDS temperature is about $14^0$C. The high temperature rise at the end of the structure is due to dipole detuning which may be reduced in the future design.

## 6 SUMMARY

The RDDS1 X-band structure is the first structure that was optimized and designed using parallel finite element code Omega3P. The Omega3P results were compared with cold test measurements. The accuracy of Omega3P predictions on the RDDS cell dimensions is better than $10^{-4}$, or 1 MHz at X-band frequency. The final machine table of the RDDS1 structure was determined numerically using Omega3P. This approach becomes possible only with advanced field solver and supercomputing resources. Omega3P is shown to be a powerful and essential tool for modeling complex rf components with unprecedented accuracy and speed.

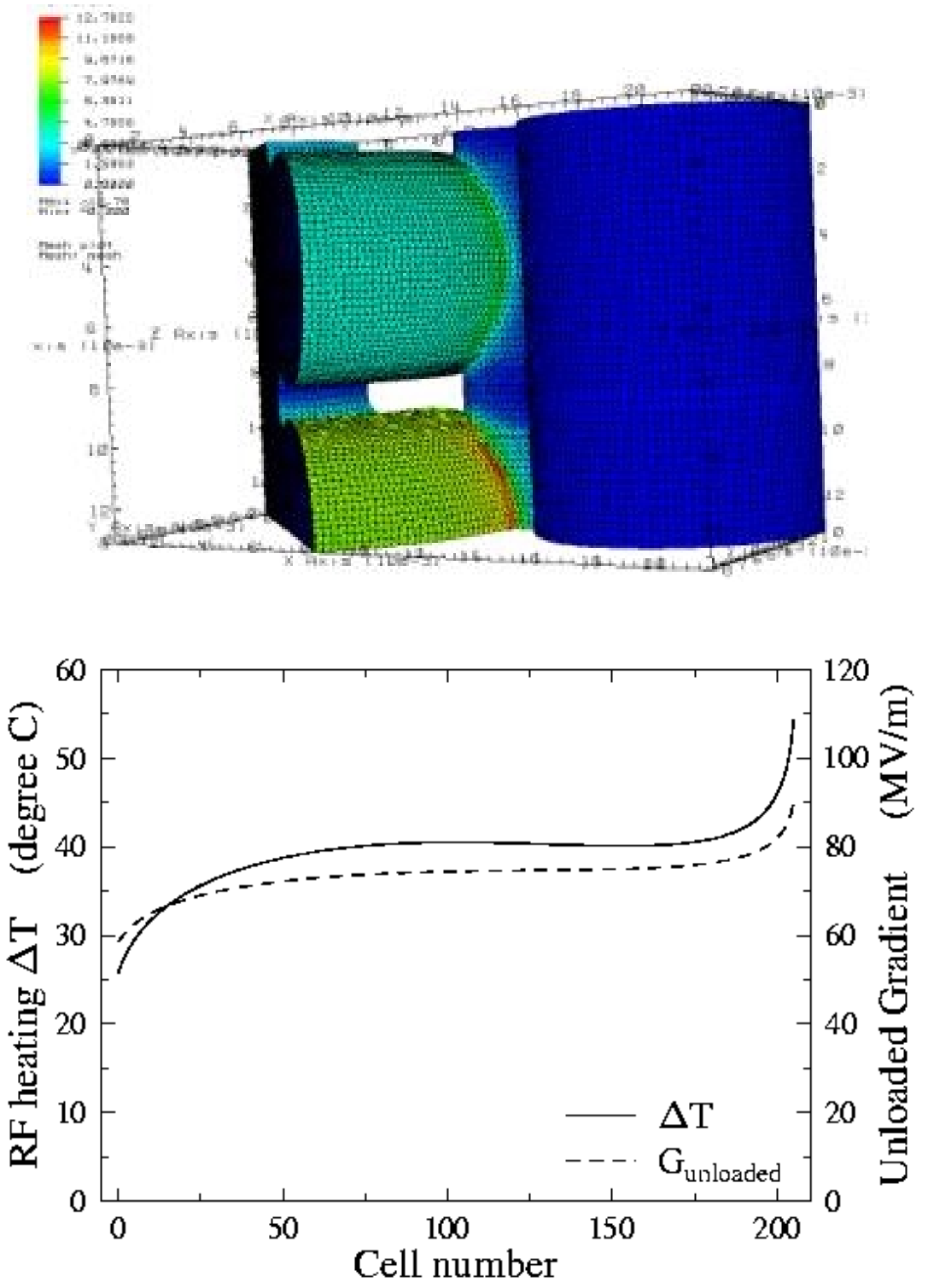

Figure 4 Pulse heating of the RDDS1 structure. Top) wall loss power distribution, clear enhancement around the wide slot; bottom) temperature rise along the structure.